\documentclass[conference,twocolumn,10pt]{IEEEtran}
\usepackage{graphicx}
\usepackage{amsmath}
\usepackage[utf8]{inputenc}
\usepackage{color}

\usepackage{graphicx}
\usepackage{rotating}
\usepackage{tikz}
\usepackage{amssymb}
\usepackage{epsfig}
\usepackage{subfigure}
\usepackage{epstopdf}
\usepackage{algorithm}
\usepackage{algorithmic}
\usepackage{amsmath}
\usepackage{amsfonts}
\usepackage{dsfont}
\usepackage{url}
\usepackage{chemarr}
\usepackage{chemarrow}
\usepackage{bbold}
\usepackage[version=3]{mhchem}
\usepackage{mathabx}
\usepackage{tabularx,booktabs}
\usepackage{booktabs}

\ifCLASSINFOpdf
\else
\fi
\IEEEoverridecommandlockouts

\begin{document}
%
\title{Characterizing Information Propagation in Plants}
%
%
%

\author{Hamdan Awan$^{\dagger *}$, Raviraj S. Adve$^\diamondsuit$, Nigel Wallbridge$^\ddagger$, Carrol Plummer$^\ddagger$, and Andrew W. Eckford$^\dagger$\\
$^\dagger$Dept. of EECS, York University, Toronto, Ontario, Canada\\
$^\diamondsuit$The Edward S. Rogers Sr. Dept. of ECE, University of Toronto, Toronto, Ontario, Canada\\
$^\ddagger$Vivent SaRL, Crans-pr\`es-C\'eligny, Switzerland\\
$^*$Corresponding author, email: hawan@eecs.yorku.ca
%
}

%
%

\markboth{IEEE Transactions~Vol.~xx, No.~xx, March~2018}%
{Shell \MakeLowercase{\textit{et al.}}: Title : Later}
%



\maketitle

\begin{abstract}
This paper considers an electro-chemical based communication model for intercellular communication in plants. Many plants, such as {\em Mimosa pudica} (the ``sensitive plant''), employ electrochemical signals known as action potentials (APs) for communication purposes. In this paper we present a simple model for action potential generation. We make use of the concepts from molecular communication to explain the underlying process of information transfer in a plant. Using the information-theoretic analysis, we compute the mutual information between the input and output in this work. The key aim is to study the variations in the information propagation speed for varying number of plant cells for one simple case. Furthermore we study the impact of the AP signal on the mutual information and information propagation speed. We aim to explore further that how the growth rate in plants can impact the information transfer rate and vice versa.
\end{abstract}


%
\IEEEpeerreviewmaketitle

\section{Introduction}

\label{sec:intro}

Recent work in biological literature suggests that electrical and electromagnetic communication in higher organisms is worth investigating. Action potentials (APs) are electrochemical signals in biological communication systems. Though commonly associated with the firing of neurons, APs also play a significant role in plants. For example, {\em Mimosa pudica}, the ``sensitive plant'', closes its leaves when touched: the signal to close the leaves is carried by an AP, as proposed by Bose over a century ago \cite{bose1914}. AP signals in plant can be defined as a sudden change or increase in the resting potential of the cell as a result of an external stimulus \cite{sukhov2009mathematical}. Some mathematical models for AP generation are presented in literature such as \cite{sukhov2009mathematical,sukhov2011simulation}. 

In this work we focus on the electrical  AP signal generation in plants and its impact on information propagation through chemical molecules. In this work we present a simple general model of an AP generation in plants. It is clear from the models \cite{sukhov2009mathematical,sukhov2011simulation,novikova2017mathematical} that understanding of APs is informed by molecular communication \cite{nakano2013-book,farsad2016comprehensive}, a communication paradigm inspired by the communication between living cells \cite{Akyildiz:2008vt,Hiyama:2010jf,Nakano:2014fq}. In this paper, we consider diffusion-based molecular communication of signalling molecules (as a result of AP signal) through a fluid medium. We will focus on the mutual information where the \textcolor{black}{receiver is based on chemical reactions i.e. ligand-receptor binding.} Furthermore in this paper we compute the mutual information between the input (number of signaling molecules) and the output number of molecules produced \textcolor{black}{by a number of receiver cells in series. }

We make two main contributions. First we study the impact of growth rate on the information propagation speed in the system. The information propagation speed can be defined as a measure of how fast the information propagates from transmitting to receiving cells. To the best of our knowledge there is a limited study about the impact of increasing length of the chain of cells on the information propagation speed. In this paper we use the mutual information for different number of receiver cells in series and  compute the information propagation speed by selecting a suitable threshold. We show that, in general, an increase in the number of cells in the chain results in an increase in information propagation speed. \textcolor{black}{Secondly we study the impact of AP signal on the mutual information and information propagation speed by comparing with the case when we have no AP signal.}

This paper is organized as follows. We describe the system model in Section \ref{system1}. We present transmitter, action potential generation and voxel model for propagation in subsection \ref{system}. Next we present the diffusion only system in subection \ref{diffusion}. This is followed by the modelling of the receiver in subsection \ref{receiver}. Section \ref{complete} presents a model for the complete system. The expressions for mutual information and the information propagation speed are derived in Section \ref{mutual}. \textcolor{black}{ Next in Section \ref{numerical} we present the results for mutual information, the information propagation speed for varying number of receivers in series and the impact of AP signal on mutual information and propagation speed.} Section \ref{conclusion} presents the conclusion.

\section{System Model}
\label{system1}
\begin{figure}
\begin{center}
\includegraphics[trim=0cm 0cm 0cm 0cm ,clip=true, width=3in]{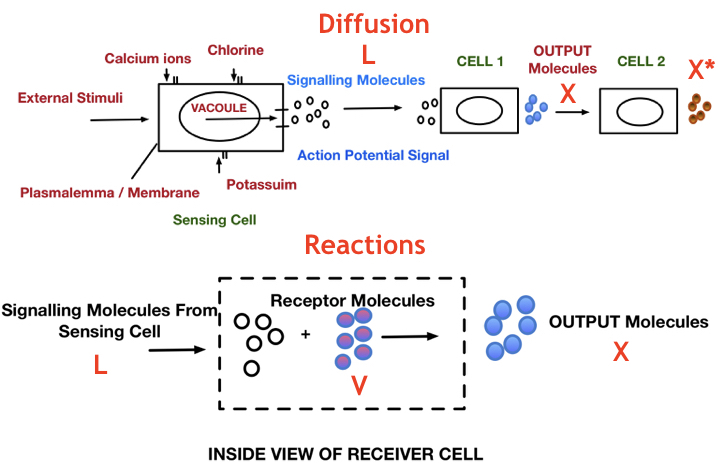}
\caption{System Model}
\label{system series}
\end{center}
\end{figure}

In this work we consider a communication link which consists of a sensing/transmitter cell and a number of receiver cells in one series as shown in Figure \ref{system series}. In the transmitter cell the AP signal is generated due to an external stimulus such as a change in temperature or electrical signal. \textcolor{black}{ As a result of this stimulus the transmitter cell emits an increased number of signalling molecules (as compared to the absence of AP signal) which diffuse freely in the propagation medium. The increased number of signalling molecules emitted by the transmitter cell is proportional to the action potential signal strength. } The signalling molecules propagate through the medium to the receiver cells where they react with the receptors to produce output molecules. The number of output molecules in the receiver over time is the output signal of the communication link. Our aim is to use the mutual information of this communication link and use this to obtain the information propagation speed.
  
\subsection{Transmitter, Action Potential and Voxel Model} 
\label{system}

\subsubsection{Sensing/Transmitter Cell}
In a typical plant cell, there is a potential difference across the cell membrane known as resting potential. \textcolor{black}{  The generation of AP is associated with passive fluxes of ions in the cell. As as a result of external factors such as an electrical signal or change in temperature, the resting potential increases to certain threshold causing ion-channels in the outer cell membrane to open up and resulting in a flow of ions into the cell. The typical ions are calcium (Ca$^{+2}$), chlorine (Cl$^-$) and potassium (K$^+$). This results in increasing the resting potential value of the cell membrane. The AP signal is generated when this resting potential crosses a certain threshold value. } \textcolor{black}{Once this AP signal is generated in the system it triggers the release of additional signalling molecules from the transmitting cell. } 


%
%

\subsubsection{Action Potential Generation: Simple Model}

Let $E_R$ represent the resting potential of the cell.  The new membrane potential $E_m$ as a result of external stimulus causing the change in ion-concentrations is \textcolor{black}{ given in \cite{sukhov2009mathematical,sukhov2011simulation} as:}
\begin{equation}
E_m= \frac{g_k E_k + g_{cl} E_{cl} + g_{ca} E_{ca}}{g_k + g_{cl}+ g_{ca}}
\label{1:sa}
\end{equation}
 \begin{equation}
\quad \textrm{where,} \quad  g_i = \frac{Fh_i}{E_R-E_i}
\end{equation}

\textcolor{black}{ Note that for simplicity we use this equation to represent the change in resting potential $E_m$ (of the transmitting/sensing cell) when the ion channels are open. A detailed discussion about the AP generation and its mathematical model will be presented in a journal version of this work.}
%
%
$g_i$ represents the electrical conductivity and $E_i$ represents the resting potential value for ion channel $i$, i.e. $E_k$ and $g_k$ for potassium ($K$) channel etc. Furthermore the term $F$ represents Faraday's constant and \textcolor{black}{$h_i$}, the ion flow across the membrane which is given as:
\begin{equation}
h_i = z  \mu P_m p_o \frac{\phi _i \eta_o - \phi_o \eta_i (\exp (-z \mu))}{1-\exp (-z \mu)} 
\end{equation}
where $z$ is ion charge. The terms $\phi _i$ and  $\phi_o $ (respectively $\eta_i$ and  $\eta_o$) represent the probability that ion is (respectively not) linked to the channel inside and outside. $P_m$ represents the maximum permeability of the cell membrane. The term $\mu$ denotes the normalized resting potential and is given as:
\begin{equation}
\mu = \frac{E_m F}{RT}
\end{equation}
where $R$ is the gas constant, and $T$ is temperature. The term $p_o$ represents the ion-channel opening state probability. For $k_1$ (channel opening)  and $k_2$ (channel closing) reaction rate constants we obtain this as:
\begin{equation}
\frac{dp_o}{dt} = k_1 (1- p_o) - k_2 (p_o)
\end{equation}

The values of all these parameters are presented in Table \ref{table:1}. \textcolor{black}{The input of the system $U(t)$ i.e. the number of signalling molecules emitted by the transmitter/sensing cell is given as:}
\begin{equation}
U(t) \; \propto \; E_m
\label{1:ua}
\end{equation}

\textcolor{black}{Where $E_m$ is given by Equation \eqref{1:sa}. Note that this $U(t)$ acts as the system input in Section \ref{complete}. This relation means that in the event of an AP signal generation the transmitter emitts higher number of molecules as compared to no AP signal. To explain this we refer to Figure \ref{system series}. Let us first assume the case when no AP signal is generated. In this case the number of molecules released at the cell membrane (to the neighbour cells) will depend only on the resting potential. We assume this as $5$ molecules per second. However as the external stimulus generates an AP signal, the number of molecules released can increase upto $20$ per second as the resting potential increases.}


%
%
\subsubsection{\textcolor{black}{ Voxel Model for Propagation}}

In this section we explain the voxel model for propagation of the signalling molecules from the sensing/transmitting cell to the receiver cell(s). The propagation occurs through the vascular bundles connecting different cells. We assume the medium of propagation is a three dimensional space of dimension $\ell_X \times \ell_Y \times \ell_Z$ where each dimension is an integral multiple of length $ \Delta$ i.e. $\ell_X = M_x \Delta$, $\ell_Y = M_y\Delta$ and $\ell_Z = M_z \Delta $. The medium is divided into $M_x \times M_y \times M_z$ cubic voxels where the volume of each voxel is $ \Delta^3$ and it represents a single cell.

\begin{figure}
\begin{center}
\includegraphics[trim=0cm 0cm 0cm 0cm ,clip=true, width=0.7\columnwidth]{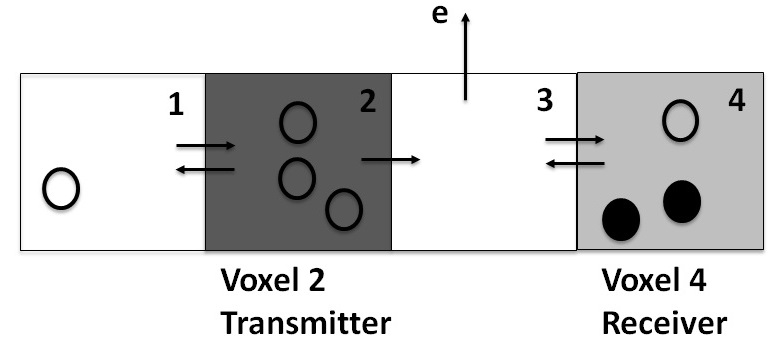}
\caption{Propagation Medium}
\label{1b}
\end{center}
\end{figure}

Figure \ref{1b} shows an example with $M_x$ = 4, $M_y$ =1 and $M_z$ = 1. For the ease of presentation we describe this 1-dimensional example. The transmitter and each receiver cell occupy a single voxel. The transmitter and receiver are assumed to be located, respectively, at the voxels with indices $T_l=2$ and $R_l=4$. The empty circles represent signaling or input molecules whereas the filled circles represent the output molecules. Diffusion is modelled as molecules moving from one voxel to a neighbouring voxel as shown by arrows in Figure \ref{1b}. \textcolor{black}{For this example we have assumed that the molecules released by the transmitter cell at voxel 2 towards voxel 3 cannot re-enter the transmitter.} The diffusion takes place at a rate of $d = \frac{D}{\Delta^2}$ where $D$ is diffusion coefficient. This means that within an infinitesimal time $\delta t$, the probability that a molecule diffuses to a neighbouring voxel is $d \delta t$. For further details see \cite{awan2017improving}.

%
\subsection{Diffusion-Only SubSystem}
\label{diffusion}

\textcolor{black}{ In this work we take the approach of dividing the system into two sub-systems, i.e., diffusion-only subsystem and reaction-only subsystem as shown in Figure \ref{system series}. } This section explains how to model the diffusion only system. Let $n_{L,i}$ denote the number of signaling molecules in the voxel $i$. In the absence of chemical reactions, the state of the system consists of only the number of signal molecules in each voxel i.e.
 \begin{equation}
n_L (t) = [n_{L,1}(t),n_{L,2}(t),n_{L,3}(t) ,n_{L,4}(t)]^T
\label{1q:a}
\end{equation}
where the superscript $T$ in Eq. \eqref{1q:a} denotes matrix transpose. The state of the system changes when a molecule diffuses from voxel 1 to a neighboring voxel 2 at a diffusion rate $dn_{L,1}$. This event causes $n_{L,1}$ to decrease by 1 and $n_{L,2}$ to increase by 1.  We can indicate this change by using the jump vector $q_{d,1} (t) = [-1 ,1, 0, 0, 0]^T$. The state of system will be $n_L (t)+q_{d,1}$ after the occurrence of this diffusion event. We also specify the corresponding jump function $W_{d,1}(n_L (t))= dn_{L,1}$ which specifies the event rate. The molecules can escape at rate $e$. Let $J_d$ be the total number of diffusion events, then we have $J_d$ jump vectors $q_{d,j}$ and jump events $W_{d,j}(n_L (t))$ where $j = 1, ..., J_d$. Combining the jump vectors and jump rate functions of all the diffusion and escape events we obtain a matrix $H$ for the medium as follows:
\begin{align}
H =  
\left[ \begin{array}{ccccc}
-d & d & 0 & 0  \\
d & -2d & d & 0  \\
0 & 0 & -d-e & d   \\
0 & 0 & d & -d  \\
\end{array} \right] 
\label{eqn:H} 
\end{align}    

The diffusion events are stochastic and hence modeled by using stochastic differential equation (SDE) \cite{gardiner2009stochastic} as follows:
\begin{align}
\dot{n}_L(t) & = \sum_{j = 1}^{J_d} q_{d,j}W_{d,j} (n_L(t)) + \sum_{j = 1}^{J_d} q_{d,j} \sqrt{W_{d,j}(n_L(t))} \gamma_j \nonumber \\
& + {\mathds 1}_T U(t)
\label{eqn:sde:do} 
\end{align}

Note this is a form of chemical Langevin equation and is similar to our previous works see \cite{awan2017improving}. There are three terms on the right-hand side of Eq.~\eqref{eqn:sde:do}. The first term describes the deterministic dynamics.  Since all the jump rates of all the diffusion events are linear, this term can be written as a product of a matrix $H$ and the state vector $n_L(t)$. 
\begin{align}
H n_L(t) & = \sum_{j = 1}^{J_d} q_{d,j}W_{d,j} (n_L(t))  
\end{align}
The second term of Eq.~\eqref{eqn:sde:do} describes the stochastic dynamics. \textcolor{black}{ The term $\gamma_j$ is continuous-time Gaussian white noise with unit power spectral density and it is needed to correctly model the stochastic noise in the system due to diffusion. } For a more detailed explanation, the reader can refer to \cite{awan2017improving,Higham:2008dl}. The third term in Eq.~\eqref{eqn:sde:do} models the input from transmitter. \textcolor{black}{ $U(t)$ from Eq. \eqref{1:ua} denotes the transmitter emission rate at time $t$. This means, in the time interval $[t,t+\delta t)$ the transmitter emits $U(t)\delta t$ signalling molecules.}

\subsection{Reaction Only Subsystem}
\label{receiver}

%
%
In this section, we present the stochastic differential equation (SDE) governing the dynamics of a reaction-only subsystem. \textcolor{black}{This subsystem includes the reactions of incoming signaling molecules (ligands) $L$ from the transmitter with the receptors $V$ in receiver to produce output molecules $X$ as shown in Figure \ref{system series}.  The count of these output molecules over time is the output signal of the system. The reactions in the series of cells continue in the same way. However due to lack of space we present the reactions at first receiver cell only.} \textcolor{black}{Note that $n_{L,R}$ and $n_X$ denote, respectively, the number of signalling molecules in the receiver voxel and the output molecules. The scalar term $n_{L,R}$ differs from the vector $n_{L}$ which refers to the number of signalling molecules in all the voxels as shown in Equation \eqref{1q:a}}.  We use a simple receiver model (based on lineraized form of ligand receptor binding) which consists of following two linear chemical reactions:
\begin{align}
L &  \rightarrow X 		& \left[ \begin{array}{cc} -1 & 1 \end{array} \right]^T&, k_+ n_{L,R}  \label{cr:rc1}  \\
X & \rightarrow L		& \left[ \begin{array}{cc} 1 & -1 \end{array} \right]^T&, k_- n_X       \label{cr:rc2}
\end{align}

\begin{table}[]
\centering
\caption{$R_v$ Matrix for different receiver circuits}
\begin{tabular}{|c|c|}
\hline
\multicolumn{1}{|c|}{Receiver }	&	\multicolumn{1}{|c|}{$R_v$ Matrix}	\\
\hline
Receiver Reactions &    $ \begin{bmatrix}  -k_{+} & k_{-} \\ k_{+} & -k_{-} \end{bmatrix}$
\\ \hline
\end{tabular}
\label{table:1s}
\end{table}

 Each reaction is described by its chemical formula (on the left-hand side), and jump vector and jump rate (on the right-hand side). The symbols $k_+$ and $k_-$ denote the reaction rate constants. In reaction \eqref{cr:rc1} the signaling molecules react at rate ${k}_{+} n_{L,R}$ to produce output molecules. The change in number of signaling and output molecules is indicated by jump vectors. Similarly we can understand the jump vector entries for  reaction \eqref{cr:rc2}. We can model the reaction  only system using stochastic differential equations for different receiver reactions. Note that the input  is $ n_{L,R} $  i.e the number of signaling molecules in receiver. The output of this subsystem is  the number of output molecules $ n_{X} $. The state vector and SDE for the reaction only system are given as:
\begin{align}
 \tilde{n}_R(t) 
 & =  \left[ \begin{array}{c|c}
 n_{L,R}(t) & n_X(t)  
\end{array} \right]^T 
\end{align} 
\begin{align}
\dot{\tilde{n}}_R(t) & = R_v \tilde{n}_R(t) + \sum_{j = J_d+1}^{J_d + J_r} q_{r,j} \sqrt{W_{r,j}(\langle \tilde{n}_R(t) \rangle)} \gamma_j 
\label{eqn:sde:ro11} 
\end{align}

Like the modeling of the diffusion only module we use jump vectors $q_{r,j}$ and jump rates $W_{r,j}$ to model the reactions in this module. $\gamma_j$ represents the continuous white noise. The reactions are indexed from $J_d+1$ to $ J_d + J_r$ where  $J_d$ is  for the diffusion only module and $J_r$ represents the reactions in the receiver. We define the matrix $R_v$  as a  2$\times$2 matrix and its entries depend on the reactions of signaling molecules in the receiver as given in Table \ref{table:1s}. In the next section we combine the diffusion only and reaction only modules to obtain a  diffusion-reaction combined system.


\section{Diffusion-Reaction Combined System}
\label{complete}

In this section, we will combine the SDE models for the diffusion-only subsystem and the reaction-only subsystem to form the complete system model. Note that the number of signaling molecules  $n_{L,R}(t)$ appears in the state vectors $n_L(t)$ and $\tilde{n}_R(t)$ of both diffusion only and receiver only modules respectively.  Therefore, the interconnection between the diffusion-only subsystem and the output module is the number of signaling molecules in the receiver voxel, which is common. 

\begin{figure}
\begin{center}
\includegraphics[trim=0cm 0cm 0cm 0cm ,clip=true, width=0.8\columnwidth]{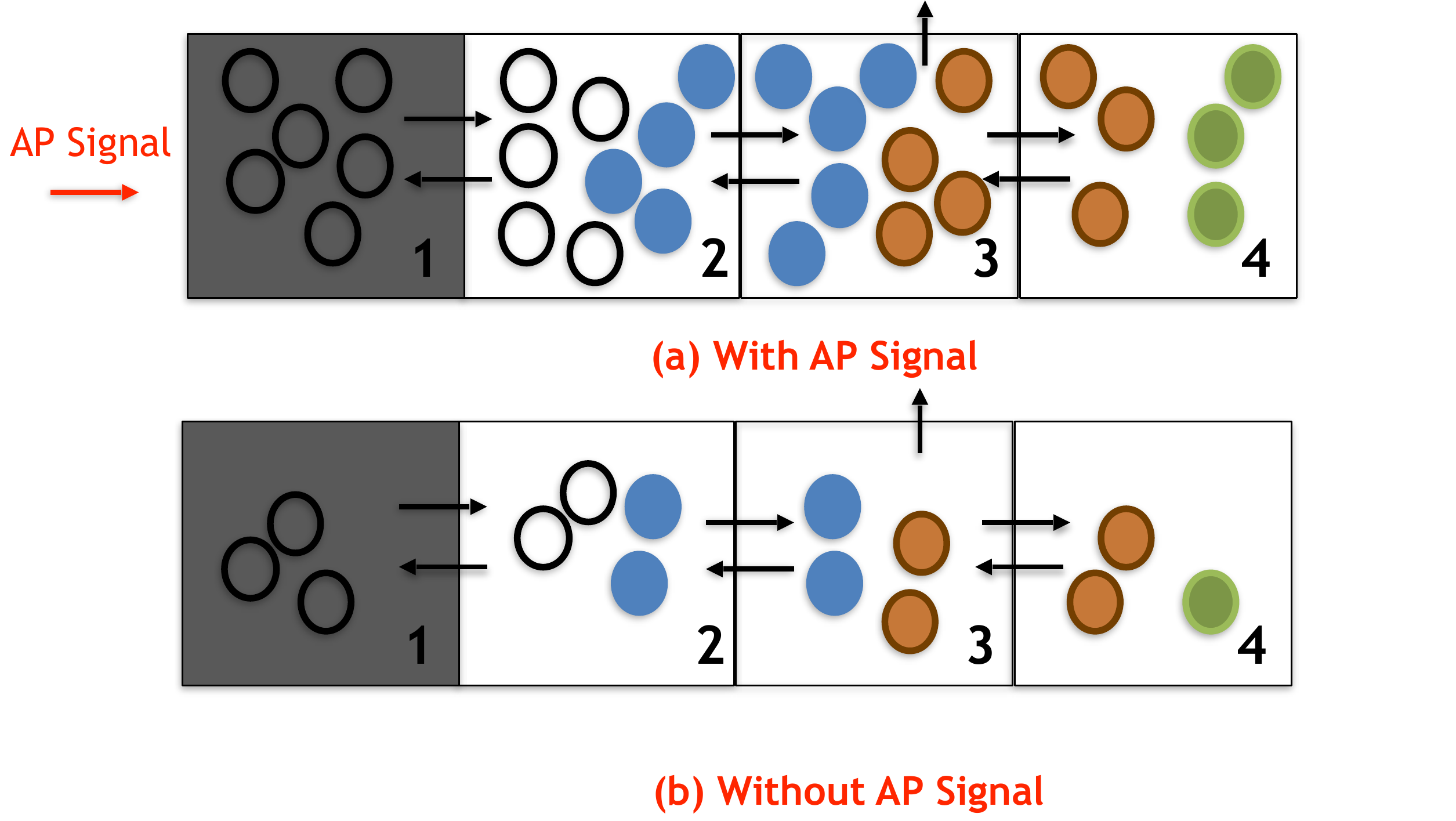}
\caption{Effect of AP Signal}
\label{AP effect}
\end{center}
\end{figure}

To illustrate our approach in this section, as an example consider the case when we have a single transmitter cell and three receiver cells; see Figure \ref{AP effect}. We compare two cases here (a) diffusion of molecules from a transmitter to receivers in the presence of an AP signal and (b) diffusion of molecules in the absence of an AP signal. As shown in the Figure the presence of an AP signal increases the input signalling molecules and hence the output number of molecules. In later sections we present the simulation results to compare the mutual information and information propagation speed for both these cases. 

We will use the example in Figure \ref{1b} to explain how the diffusion-only subsystem and the reaction only system can be combined together. We consider the dynamics of the diffusion-only subsystem for  the example, when the receiver voxel has the index $R = 4$. The evolution of the number of signaling molecules in the receiver voxel $n_{L,R}(t)$ is given by the $R$-th (i.e. fourth) row of Eq.~\eqref{eqn:sde:do} i.e.: 
\begin{align}
& \dot{n}_{L,R}(t) = d n_{L,3}(t) - d n_{L,R}(t) \nonumber \\
& + \underbrace{\sum_{j = 1}^{J_d} [q_{d,j}]_R \sqrt{W_{d,j}(n_L(t))} \gamma_j}_{\xi_d(t)}
\label{eqn:sde:ro:r2}  
\end{align}
where $[q_{d,j}]_R$ denote the $R$-th element of the vector $q_{d,j}$. The dynamics of the number of signaling molecules in the receiver voxel due to the reactions in the receiver is given by the first element of Eq.~\eqref{eqn:sde:ro11}, which is: 
\begin{align}
& \dot{n}_{L,R}(t) = R_{11} n_{L,R}(t) + R_{12} n_X(t) \nonumber \\
& + \underbrace{\sum_{j = J_d+1}^{J_d+J_r} [q_{r,j}]_1 \sqrt{W_{r,j}(\tilde{n}_R(t))} \gamma_j}_{\xi_r(t)} 
\label{eqn:sde:ro:r1a} 
\end{align}
where $[q_{r,j}]_1$ denotes the first element of the vector $q_{r,j}$. For the complete system the dynamics of $n_{L,R}(t)$ is obtained by combining Eqs. \eqref{eqn:sde:ro:r2} and \eqref{eqn:sde:ro:r1a} as follows: 
\begin{align}
\dot{n}_{L,R}(t) = &  d n_{L,3}(t) - d n_{L,R}(t)  \nonumber \\ &  + 
R_{11} n_{L,R}(t)  + R_{12} n_X(t) + \xi_{total}(t) 
\label{eqn:sde:nlr:ex} 
\end{align}
where $\xi_{total}(t) = \xi_d(t) + \xi_r(t)$.  We are now ready to describe the complete model. Let $n(t)$ be the state of the complete system and it is given by: 
\begin{align}
n(t) = 
 & \left[ \begin{array}{c|c}
 n_{L}(t)^T & n_X(t)  
\end{array} \right]^T
\label{eqn:state} 
\end{align}
We use $q_j$ and $W_j(n(t))$ to denote the jump vectors and jump rates of the combined model. The complete system SDE is:
\begin{align}
\dot{n}(t) & = A n(t) + \sum_{i = 1}^{J} q_j \sqrt{W_j(n(t))} \gamma_j + {\mathds 1}_T U(t) 
\label{eqn:mas11}
\end{align} 
where $J = J_d+J_r$, and the matrix $A$ is defined by $A n(t) = \sum_{i = 1}^{J} q_j W_j(n(t))$. \textcolor{black}{ The input $U(t)$ depends on $E_m$ as shown in Equation \eqref{1:ua}.} The matrix $A$ has the block structure: 
\begin{align}
A = 
 & \left[ \begin{array}{c|c}
H + {\mathds 1}_R^T {\mathds 1}_R R_{11}  &   {\mathds 1}_R R_{12}  \\ \hline 
R_{21}  {\mathds 1}_R^T & R_{22} 
\end{array} \right]
\label{eqn:A} 
\end{align}
where $H$ comes from the diffusion only subsystem. Similarly the $R_{ii}$ terms  come from the reaction only subsystem ($R_v$ matrix). The vector ${\mathds 1}_R$ is a unit vector with an 1 at the $R$-th position; in particular, note that  ${\mathds 1}_R^T n_L(t) = n_{L,R}(t)$ which is the number of signalling molecules in the receiver voxel. Note that, the coupling between the diffusion-only subsystem and the output module, as exemplified by Eq. \eqref{eqn:sde:nlr:ex}, takes place at the $R$-th row of $A$. Next we compute the mutual information using the Laplace transform of expression in Eq. \eqref{eqn:mas11}.

\subsection{Mutual Information and Capacity}
\label{mutual}

The input and output signals for the complete system are, respectively, the production rate $U(t)$ of the signalling molecules in the transmitter voxel (dependent on AP signal) and the number of output molecules $n_X(t)$ in the receiver voxel. In this section, we will derive an expression for the mutual information between the input $U(t)$ and output $n_X(t)$. We begin by stating a result in \cite{tostevin2010mutual} which states that, for two Gaussian distribution random processes $a(t)$ and $b(t)$, their mutual information $I(a,b)$ is: 
\begin{align}
I(a,b) &= \frac{-1}{4\pi} \int_{-\infty}^{\infty} \log \left( 1 - \frac{|\Phi_{ab}(\omega)|^2}{\Phi_{aa}(\omega) \Phi_{bb}(\omega)}  \right) d\omega
\label{eqn:MI0}
\end{align} 
where $\Phi_{aa}(\omega)$ (resp. $\Phi_{bb}(\omega)$) is the power spectral density of $a(t)$ ($b(t)$), and $\Phi_{ab}(\omega)$ is the cross spectral density of $a(t)$ and $b(t)$. In order to apply the above results to the communication link given in Eq.~\eqref{eqn:mas11}, we need a result from \cite{warren2006exact} on the power spectral density of systems consisting only of chemical reactions with linear reaction rates. Following from \cite{warren2006exact} if all the jump rates $W_j(n(t))$ in Eq. \eqref{eqn:mas11} are linear in $n(t)$, then the power spectral density of $n(t)$ is obtained by using following:
\begin{align}
\dot{n}(t) & = A n(t) + \sum_{i = 1}^{J} q_j \sqrt{W_j(\langle n(\infty) \rangle)} \gamma_j + {\mathds 1}_T U(t) 
\label{eqn:complete2}
\end{align} 
where $\langle n(t) \rangle$ denotes the mean of $n(t)$ and is the solution to the following ordinary differential equation:
\begin{align}
\dot{\langle n(t) \rangle} & = A \langle n(t) \rangle+ {\mathds 1}_T c 
\label{eqn:ode_ninfinity}
\end{align} 
where $c$ is the mean of input $U(t)$.  s a result, the dynamics of the complete system in Eq.~\eqref{eqn:complete2} are described by a set of linear SDE with $U(t)$ as the input and $n_X(t)$ (which is the last element of the state vector $n(t)$) as the output. The input $U(t)$ has the form $U(t) = c + w(t)$ where $c$ (mean of input) depends on $E_m$ and $w(t)$ is a zero-mean Gaussian random process. The noise in the output $n_X(t)$ is caused by the Gaussian white noise $\gamma_j$'s in Eq.~\eqref{eqn:complete2}. Therefore, Eq.~\eqref{eqn:complete2} models a continuous-time linear time-invariant (LTI) stochastic system subject to Gaussian input and Gaussian noise.

The power spectral density $\Phi_{X}(\omega)$ of the signal $n_X(t)$ can be obtained from standard results on the output response of a LTI system to a stationary input and is given by: 
\begin{align}
\Phi_{{X}}(\omega) & =  |\Psi(\omega) |^2 \Phi_e(\omega) + \Phi_{\eta}(\omega)
\label{2331a} 
\end{align}
where $\Phi_e(\omega)$ is the power spectral density of $E(t)$ and $|\Psi(\omega)|^2$ is the channel gain with $\Psi(\omega) = \Psi(s)|_{s = i\omega}$ defined by:
\begin{align}
\langle N_{X} (s) \rangle  & = {\mathds 1}_X \langle N(s)  \rangle = 
\underbrace{ {\mathds 1}_X  (sI - A)^{-1} {\mathds 1}_T }_{\Psi(s) }   U(s)
\label{21a}
\end{align}
Note that Eq.~\eqref{21a} can be obtained from Eq.~\eqref{eqn:complete2} after taking the mean and applying Laplace transform The transfer function $\Psi(s)$  takes into account the consumption of signaling molecules, the interaction between output molecules and the signaling molecules, as well as the possibility that a signaling molecule may leave or return in the receiver.  For details see \cite{awan2017improving}. The term $\Phi_{\eta}(\omega)$ denotes the stationary noise spectrum and is given by: 
\begin{align}
\Phi_{\eta}(\omega) & =   \sum_{j = 1}^{J_d + J_r} | {\mathds 1}_X (i \omega I - A)^{-1} q_j |^2 W_j(\langle n_{}(\infty) \rangle) 
\label{eqn:spec:noise2} 
\end{align} 
where $n_{}(t)$ denotes the state of the complete system in Eq. \eqref{eqn:state} and $\langle n_{}(\infty) \rangle$ is the mean state of system at time $\infty$ due to constant input $c$. Similarly, by using standard results on the LTI system, the cross spectral density $\Psi_{xe}(\omega)$ is:
\begin{align}
|\Psi_{xu}(\omega)|^2 &= |\Psi(\omega) |^2 \Phi_e(\omega)^2 
\label{eqn:csd} 
\end{align} 
By substituting Eq.~\eqref{2331a} and Eq.~\eqref{eqn:csd} into the mutual information expression in Eq.~\eqref{eqn:MI0}, we arrive at the mutual information $I(n_{X},U)$ between $U(t)$ and $n_{X}(t)$ is:
\begin{align}
I(n_{X},U) = \frac{1}{2} \int \log \left( 1+\frac{ | \Psi(\omega) |^2}{\Phi_{\eta}(\omega)} \Phi_e(\omega) \right) d\omega
\label{eqn:mi1}
\end{align}

The capacity or maximum mutual information of the communication link can be determined by applying the water-filling solution to Eq. \eqref{eqn:mi1} subject to a power constraint on input $U(t)$ \cite{gallager1968information}.

\subsection{Information rate vs length of cell chain}

In this section we discuss how we use of the mutual information to obtain the relationship between information propagation speed and number of cells in the chain. First we obtain the mutual information when we have number of receiver cells in series. The next step is to choose a suitable threshold value so that we can calculate the time difference at which the mutual information curve for each case crosses the threshold value. Next we use the following equation for calculating the information propagation speed $V$ (cells/sec):

\begin{equation}
  V=  \frac{1}{\mathbf{E} [\Delta t_{i,i+1}]}
\end{equation}
 
Where $\Delta t_{i,i+1}$ represents the time difference at which the mutual information for each case (i.e. increasing receivers) crosses the threshold value. \textcolor{black}{ $\mathbf{E}$ denotes the expectation operator.} This technique is used to compute the propagation speed for an increasing number of receiver cells in the chain in series. We present the results for this approach in numerical examples section.

\begin{table}[]
\centering
\caption{Parameters and their default values.}
\begin{tabular}{|c|c|}
\hline
\multicolumn{1}{|c|}{Symbols}	&	\multicolumn{1}{|c|}{Notation and Value}	\\
 \hline
$E_R$ &     Resting Potential  = -150-170 mV
\\ \hline
$F$ &    Faraday's constant = $9.65 \times 10^4 C/mol$ 
\\ \hline
$C$ &     Membrane capacity = $10^{-6} F cm^-2$  
\\ \hline
$P_m$  &     Permeability per unit area  = $10^{-6}$ M cm $s^{-1}$
\\ \hline
 $\mu$ &    $E_R$ F/ RT where T= Temperature
\\ \hline
$\gamma$ &  ratio of  rate constants = $9.9 \times 10^-5 M$
\\ \hline
 $\phi_{i}$ & Probability ion link - inside = $c_{in}$ / ($c_{in}$ + $\gamma$)
\\ \hline
 $\phi_{o}$ & Probability ion link - outside =  $c_{out}$ / ($c_{out}$ + $\gamma$)
 \\ \hline
 $\eta_i$ & Probability ion not linked-inside = 1- $\phi_{i}$ 
\\ \hline
 $\eta_o$ & Probability ion not linked-outside = 1- $\phi_{o}$
\\ \hline
$c_{in}$ and $c_{out}$ &  1.28	and 1.15 respectively
\\ \hline
$z$ & ion charge e.g. for calcium = +2.  
\\ \hline
$p_o$ & ion channel open-state probability 
\\ \hline
\end{tabular}
\label{table:1}
\end{table}


\begin{figure}
\begin{center}
\includegraphics[trim=0cm 0cm 0cm 0cm ,clip=true, width=0.9\columnwidth]{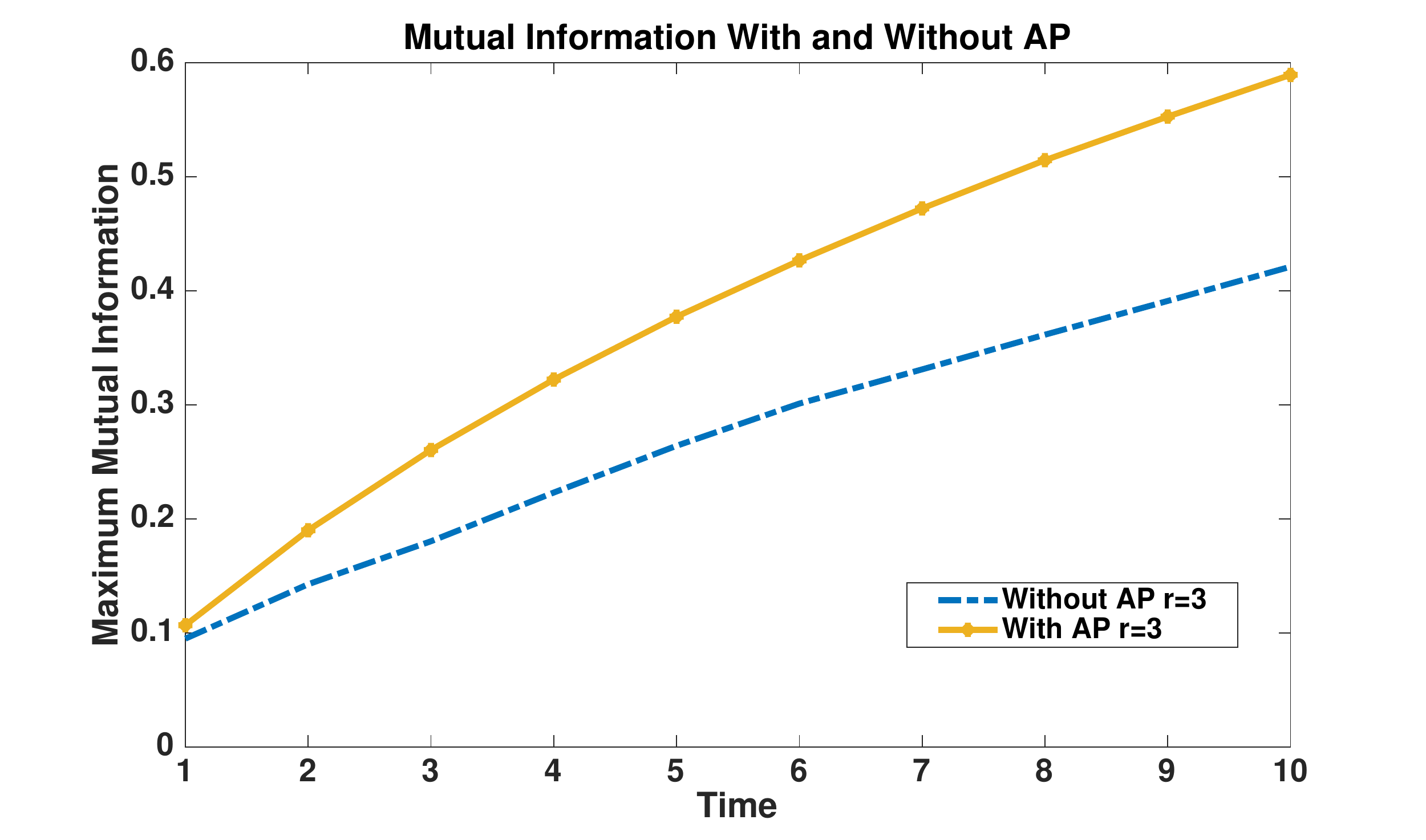}
\caption{\textcolor{black}{Mutual Information With and Without AP}}
\label{Ap-m}
\end{center}
\end{figure}

\begin{figure}
\begin{center}
\includegraphics[trim=0cm 0cm 0cm 0cm ,clip=true, width=0.9\columnwidth]{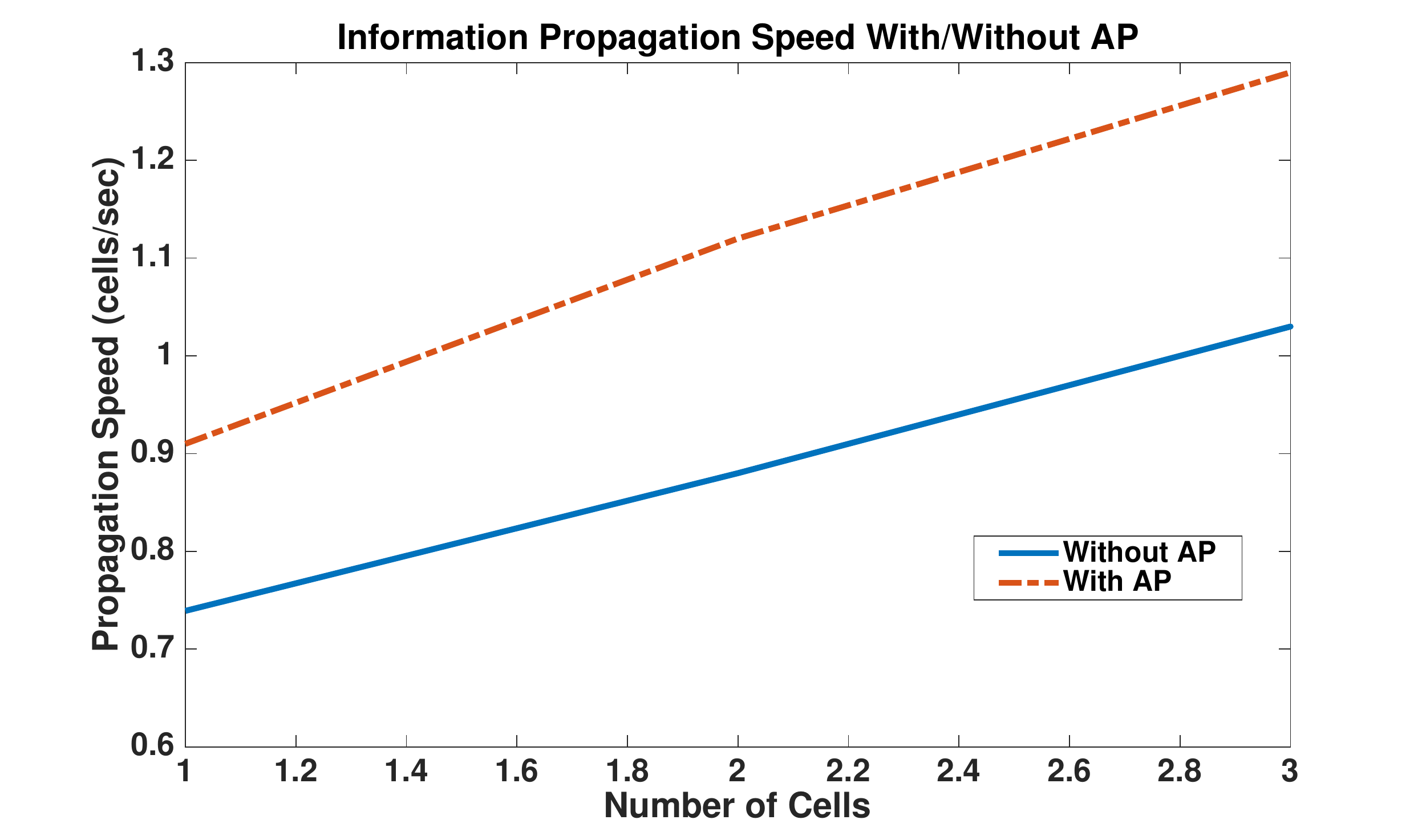}
\caption{\textcolor{black}{Information Propagation Speed With and Without AP-Series Case}}
\label{Ap-ip}
\end{center}
\end{figure}

\section{Numerical Examples- Simulations}
\label{numerical}
 
In this section we discuss the numerical results related to this work. The parameters used for the generation of AP signal are given in Table \ref{table:1}. The magnitude of the generated AP signal can be typically in the range of 20-80 mV. The AP signal generation results are not included due to limited space. We present these results in journal extension of this paper. For this system we obtain an action potential signal of about 60 millivolts which will trigger the release of signaling molecules from the transmitting cell to the receptor cell(s). For propagation medium we assume a voxel size of ($\frac{1}{3}$$\mu$m)$^{3}$ (i.e., $\Delta = \frac{1}{3}$ $\mu$m), creating an array of $4\times 1 \times 1$ voxels for the series configuration. The transmitter and each receiver occupy one voxel each as mentioned in the system model in Section \ref{system}. We assume the diffusion coefficient $D$ of the medium is $1$ $\mu$m$^2$s$^{-1}$. The mean emission rate $c$ is dependent on the AP signal which triggers release of molecules. The aim is to compute the mutual information between the input and output number of molecules of the complete system for number of receivers in series and use that to study information propagation speed for increasing number of cells.




For this paper we show the result for the case with single sensing/transmitting cell and three receiver cells in series. \textcolor{black}{In this work we present the result where we show the impact of AP signal on the mutual information and information propagation speed. In Figure \ref{Ap-m} we show that the mutual information  increases in the presence of AP signal for the system with single transmitter and three receiver cells as shown in Figure \ref{AP effect}. Next by using the mutual information and selecting a threshold value we show that the information propagation speed increases in the presence of an AP signal as shown in Figure \ref{Ap-ip}. Another important result from Figure \ref{Ap-ip} is that the information propagation speed increases with the increase in the number of receiver cells in series.}

\textcolor{black}{The results for the mutual information as well as information propagation speed for the parallel receiver case will be discussed in the journal extension of this paper. } 







\section{Conclusion}
\label{conclusion}

In this paper we presented a simple model for the generation of action potential signal in plants. We realize the information transfer from a transmitter cell to a number of receiver cells in series and computed the mutual information between the input signal from transmitting cell and output signal of the receiver cell(s).  By using the values of mutual information and selecting a threshold we obtained the information propagation speed as a function of the number of cells in the chain. We realize that  the information propagation speed tends to increase with an increasing number of receiver cells in series. \textcolor{black}{We further show that the presence of an AP signal leads to an increase in mutual information and information propagation speed for an increasing number of receiver cells.}



%





\ifCLASSOPTIONcaptionsoff
  \newpage
\fi

\bibliographystyle{ieeetr}
\bibliography{nano2017,book,nano2018}

\end{document}